\documentclass[prl,twocolumn,epsfig,rotate,showpacs]{revtex4}
\usepackage{epsfig}
\usepackage{graphicx}
\usepackage{amsmath}

\newcommand{\beq}{\begin{eqnarray}}
\newcommand{\eeq}{\end{eqnarray}}
\begin{document}

\title{Butterfly Floquet Spectrum in Driven SU(2) Systems}
\author{Jiao Wang$^{1,2}$ and Jiangbin Gong$^{3,4}$}
\email{phygj@nus.edu.sg} \affiliation{$^{1}$Temasek Laboratories,
National University of Singapore, 117542, Singapore
\\ $^{2}$Department of Physics, Institute of Theoretical Physics and Astrophysics, Xiamen University, Xiamen 361005, China
\\ $^{3}$Department of Physics and Center of Computational Science and Engineering,
National University of Singapore, 117542, Singapore
\\ $^{4}$NUS Graduate School for Integrative Sciences and Engineering, Singapore
 117597, Singapore}
\date{Dec. 29, 2008}

\begin{abstract}
The Floquet spectrum of a class of driven SU(2) systems is shown to
display a butterfly pattern with multi-fractal properties.  The
level crossing between Floquet states of the same parity or
different parities is studied. The results are relevant to studies
of fractal statistics, quantum chaos, coherent destruction of
tunneling, and the validity of mean-field
descriptions of Bose-Einstein condensates. %This work will also motivate
%interests in engineering and simulation of driven systems with
%intriguing spectral properties.
\end{abstract}

\pacs{03.75.Lm, 05.45.Mt, 03.75.-b} \maketitle

Hofstadter's butterfly spectrum of the Harper model \cite{hof} has
attracted tremendous mathematical, theoretical and experimental
interests.  For an arbitrary irrational value of one system
parameter, the spectrum of the Harper model is a fractal, which
has been strictly proved after decades of research on the ``Ten
Martini problem" \cite{simon}.  As one important implication, a
fractal butterfly spectrum suggests the closing of a quantum gap
infinite times and hence the occurrence of infinite quantum phase
transitions \cite{goldman}.

Early quantum chaos studies established that the Floquet
(quasi-energy) spectrum of periodically driven systems may display a
fractal butterfly pattern as well \cite{kickedharper,jiao08}.
However, the nature of the fractal Floquet spectrum is still poorly
understood. Indeed, because the eigen-phase of Floquet states is
restricted to a range of $2\pi$, understanding a Floquet spectrum
associated with an infinite-dimensional Hilbert space is subtle and
challenging \cite{kohn}. Furthermore, previous findings regarding to
the fractal Floquet spectrum were largely limited to the so-called
kicked-Harper model (a driven version of the Harper model)
\cite{casatietal,dana1,dana2} and its variant \cite{jiao08,lawton,dana2}. % But even
%in such few cases a rigorous mathematical proof about the fractal
%nature of a butterfly Floquet spectrum is still lacking
%\cite{lawton}.

Due to vast interests in quantum control especially in dressed
matter-waves \cite{holthaus,arimondo,Korsch,Qi}, there are now
promising possibilities for the engineering and simulation of driven
ultracold systems with a prescribed Floquet spectrum.  In this
Letter, we show that the Floquet spectrum of a deceptively simple
class of SU(2) systems, constructed from a driven two-mode
Bose-Einstein condensate (BEC), displays a butterfly pattern and
possesses truly remarkable properties.  For example, we show that
with one certain system parameter fixed the overall butterfly
pattern is insensitive to the number of bosons (denoted $N$) in the
BEC, but some detailed features depend on whether $N$ is odd or
even. We shall reveal that the found butterfly pattern contains many
level crossings between states of different parities and thus many
points of coherent destruction of tunneling (CDT) \cite{Hanggi},
with the total number of CDT points found to scale as $\sim
N^{3.0}$. As an analog of first-order quantum phase transitions, we
discover that the found butterfly pattern also contains many level
crossings between same-parity eigenstates. These results suggest
that the class of driven SU(2) systems studied here may become a
test bed for a number of research topics. Several specific
applications of this work are also discussed.

Driven two-mode BEC systems were proposed before
\cite{Milburn,Korsch,liujie} to realize the well-known kicked top
model \cite{Haakebook} in the quantum chaos literature. In its most
general form, a driven two-mode Bose-Hubbard model can be written as
\begin{eqnarray}
H=f(t)\hbar (a_{1}^{\dagger}a_2+ a_{2}^{\dagger}a_1)+ g(t)\hbar
(a_{1}^{\dagger}a_1-a_{2}^{\dagger}a_2)^2, \label{Hami}
\end{eqnarray}
where $a_{i}$ and $a_{i}^{\dagger}$ are the bosonic annihilation and
creation operators for the $i$th mode, $f(t)$ describes the
time-dependent tunneling rate between the two modes, and the $g(t)$
term describes the self-interaction between same-site bosons, whose
time-dependence can be achieved by Feshbach resonance induced by a
magnetic field.  The total number of bosons
$N=a_{1}^{\dagger}a_1+a_2^{\dagger}a_2$ is a conserved quantity and
the dimension of the Hilbert space is $N+1$.  Using the Schwinger
representation of angular
 momentum operators, namely, $J_{x}=
(a_{1}^{\dagger}a_2+a_{2}^{\dagger}a_1)/2$,
 $J_y=(a_{2}^{\dagger}a_1-a_{1}^{\dagger}a_2)/(2i)$, and
 $J_z=(a_{1}^{\dagger}a_1-a_{2}^{\dagger}a_2)/2$, Eq. (\ref{Hami})
 reduces to
 \begin{eqnarray}
H= 2f(t)\hbar J_x+ 4g(t)\hbar J_z^2. \label{Hami2}
\end{eqnarray}
Clearly, the dynamics is solely determined by the SU(2) generators
$J_x$, $J_y$ and $J_z$. The total angular momentum quantum number
$J$ is given by $J=N/2$. The Hilbert space can be expanded by the
eigenstates of $J_z$, denoted $|m\rangle$, with
$J_z|m\rangle=m|m\rangle$.  The population difference between the
two modes is given by the expectation value of $2J_{z}$.  If we
exchange the indices of the two modes, then $J_x$ is invariant,
$J_z\rightarrow -J_z$, and as a result the Hamiltonian in Eq.
(\ref{Hami2}) is unchanged. This reflects a parity symmetry of our
model.

Consider then two specific forms of $f(t)$ and $g(t)$. In the first
case $f(t)=\alpha/(2\tau)$, $g(t)= g_0
\sum_{n}[\delta(t-2n\tau-\tau)-\delta(t-2n\tau)]$. The Floquet
operator, i.e., the unitary evolution operator $F$ from
$2n\tau+0^{+}$ to $(2n+2)\tau+0^{+}$, is then given by
\begin{eqnarray}
F=e^{i\eta J_z^2/2J}e^{-i \alpha J_x} e^{-i\eta J_z^2/2J}
e^{-i\alpha J_x}, \label{Feq}
\end{eqnarray}
where $\eta=4 g_0 N$.  Because the first two or the last two factors
in Eq. (\ref{Feq}) constitute the Floquet operator for a standard
kicked-top model \cite{Haakebook}, our driven system here can be
regarded as a ``double-kicked top model". Alternatively, if we set
$g(t)=g_0/\xi$,
$f(t)=\frac{\alpha}{2}\sum_{n}[\delta(t-n\tau)+\delta(t-n\tau-\xi)]$,
where $\xi$ is the time delay between the two delta kicking
sequences, then the associated propagator $F'$ from $n\tau-0^{+}$ to
$(n+1)\tau-0^{+}$ is given by
\begin{eqnarray}
F'=e^{i\eta J_z^2/2J}e^{-i(4g_0\tau/\xi)J_z^2}e^{-i\alpha J_x}
e^{-i\eta J_z^2/2J} e^{-i\alpha J_x}. \label{2ndF}
\end{eqnarray}
Under the special condition $4g_0\tau/\xi=2k\pi$ ($8k\pi$) for
integer $J$ (half integer $J$), where $k$ is an integer , the factor
$e^{-i(4g_0\tau/\xi) J_z^2}$ is unity in the $(2J+1)$-dimensional
Hilbert space and hence $F'$ becomes identical with $F$.  Thus,
there exist two different scenarios for realizing $F$, the key
operator to be analyzed below.

In the $|m\rangle$ representation, the third factor $e^{-i\eta
J_z^2/2J}$ of $F$ equals $e^{-i\eta m^2/2J}$, which is a
pseudo-random number for irrational $\eta/J$. The first factor of
$F$ however effectively induces a time-reversal of the third
factor and thus partially cancels this pseudo-random phase.
Indeed, using the SU(2) algebra, the product of the first three
factors of $F$ in Eq. (\ref{Feq}) is given by
\begin{eqnarray}
& & e^{i\eta J_z^2/2J}e^{-i \alpha J_x} e^{-i\eta J_z^2/2J}\nonumber
\\
&=& e^{-i\alpha\left \{(J_x/2+iJ_y/2)e^{i[\eta (2J_z+1)/2J]} +
\text{c.c.}\right\}},
\end{eqnarray}
showing that the $\eta$-dependent term entering into $F$ becomes
$e^{i[\eta (2J_z+1)/2J]}$, which is always a quasi-periodic number
$e^{i[\eta (2m+1)/2J]}$ in the $|m\rangle$ representation. This
partial cancelation of quasi-random dynamical phases implies
intriguing spectral properties \cite{gong07}.

To study the classical limit of $F$ we consider scaled variables
$x=J_x/J$, $y=J_y/J$, and $z=J_z/J$. The three operators $x$, $y$,
$z$ also satisfy the angular-momentum algebra, but with an effective
Planck constant $\hbar_{\text{eff}}\equiv 1/J$. Taking the
$\hbar_{\text{eff}} \rightarrow 0$ limit with fixed $\eta$ and
$\alpha$, the classical dynamics associated with $F$ can be
obtained, with variables $x$, $y$, and $z$ restricted on a unit
sphere. Because $\eta=4g_0 N$, this classical limit with fixed
$\eta$ requires $N\rightarrow +\infty$ and $g_0\rightarrow 0$. This
condition is apparently equivalent to that in a standard mean-field
limit of the driven BEC.

%**********************************************************************************fig1
\begin{figure}
\vspace{-.2cm}\hspace{0.cm}\epsfig{file=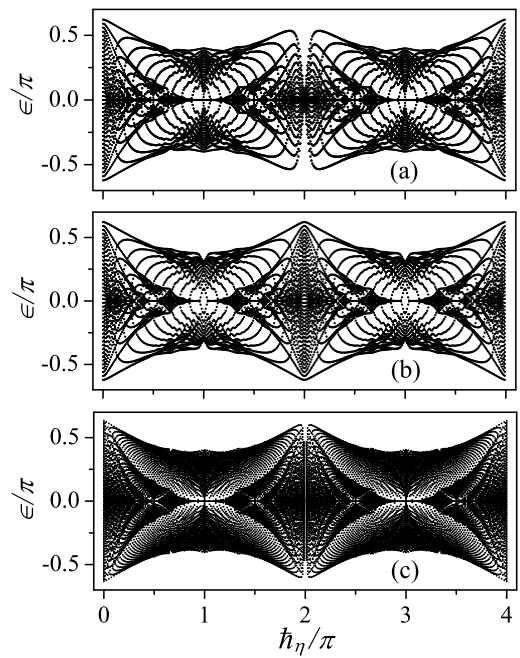,width=7.5cm}
\vspace{-1.7cm}\caption{Eigen-phase spectrum (denoted $\epsilon$) of
the Floquet operator $F$ in Eq. (3). $J=20$ in (a), 20.5 in (b), and
100 in (c). $\alpha/\hbar_{\text{eff}}=\alpha\cdot J=1.0$ in all
panels. }\label{fig1}
\end{figure}

Figure 1 shows the typical eigen-phase spectrum of $F$ vs
$\hbar_{\eta}\equiv \eta \hbar_{\text{eff}}=\eta/J=8g_0$, for $J=20,
20.5, 100$ and $\alpha/\hbar_{\text{eff}}=1.0$.  Because the
spectrum of $F$ is invariant if $\hbar_{\eta}\rightarrow
\hbar_{\eta}+4\pi$, we set $\hbar_{\eta}\in[0, 4\pi)$. Though in
Fig. 1 the involved Hilbert space is rather small, spectacular
butterfly patterns are already obtained (their symmetry with respect
to $\hbar_\eta=2\pi$ can be proved). They resemble the famous
Hofstadter's butterfly, but also present remarkable differences in
several aspects. First, if we take a vertical cut of the butterfly
patterns in Fig. 1, the spectrum is not found to present any large
gap. Second, the butterfly patterns shown in each panel of Fig. 1
possess a double-butterfly structure, with each butterfly covering a
$2\pi$ range of $\hbar_{\eta}$. This double-butterfly structure is
somewhat analogous to the spectrum of a Harper-like effective
Hamiltonian considered in Ref. \cite{dana2}.  More interestingly,
though Fig. 1(c) has much more levels than Fig. 1(a) and Fig. 1(b),
the overall outline of the double-butterfly structure is seen to be
insensitive to $J$ for fixed $\alpha/\hbar_{\text{eff}}=\alpha\cdot
J$.  For a fixed value of $J$ but for other not too large values of
$\alpha$, the qualitative features of the butterfly spectrum remain,
but at different scales. For very large values of $\alpha$ (e.g.,
$\alpha/\hbar_{\text{eff}}>10$), the butterfly pattern for a fixed
value of $J$ will gradually dissolve, in a similar fashion as in the
kicked Harper model \cite{kickedharper}.

Some detailed aspects of the spectrum are also noteworthy.  For
example, it is observed that the spectrum collapses to one point
for $\hbar_{\eta}=2\pi$, if and only if $J$ is an integer. This
can be explained as follows. If $J$ is an integer and if
$\hbar_{\eta}=2\pi$, then in the $|m\rangle$ representation,
$e^{-i\eta J_z^2/2J}=e^{-i\pi m^2}=e^{-i\pi m}=e^{-i\pi J_z}$. So
in this case $e^{-i\eta J_z^2/2J}$ is equivalent to a rotation of
$\pi$ around the $z$ axis, and hence the first three factors of
$F$ exactly cancel its last factor. This cancelation will not
occur if $J$ is a half integer, i.e., if $N$ is odd.

Figure 2 depicts the phase space structure of the classical limit of
$F$. As $\eta$ increases, the classical dynamics changes from being
regular to being chaotic. On the other hand,  the Floquet spectrum
shown in Fig. 1 can be however much similar for radically different
values of $\eta$ ranging from 0 to $4\pi/\hbar_{\text{eff}}=4J\pi$.
Therefore, upon quantization the regular or chaotic nature of the
classical dynamics might not necessarily be reflected in the
spectrum and hence can be irrelevant to the quantum dynamics.

%**********************************************************************************fig2
\begin{figure}
\vspace{0.3cm}\hspace{0.15cm}\epsfig{file=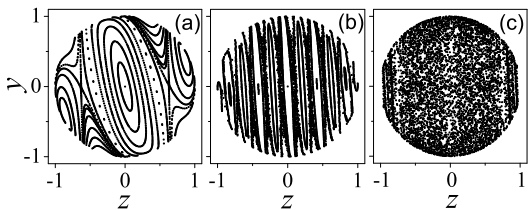,width=7.2cm}
\caption{Poincar\'{e} surfaces of section (with $J_x>0$) of the
classical or mean-field limit of $F$ in Eq. (3), with $\alpha=0.05$
(same as in Fig. 1(a)), $\eta=5$ in (a), 30 in (b), and 75 in
(c).}\label{fig2}
\end{figure}

The statistical behavior of the found butterfly spectrum is also
examined. To have good statistics we consider a much larger value of
$J$. Figure 3(a) presents the cumulative level density $N(\epsilon)$
for a representative value of $\hbar_{\eta}$. $N(\epsilon)$ is
highly irregular, but does not show any clear flat steps. This is
consistent with our early observation that no large gap exists in
the spectrum. Figure 3(b)-3(d) shows the associated level
distribution $P(\epsilon)$ at three different scales. Evidently,
$P(\epsilon)$ has a fascinating self-similar property. This
motivates us to quantitatively characterize the spectrum via the
generalized fractal dimension $D_{q}$, with the results shown in
Fig. 3(e). As expected from the $N(\epsilon)$ result in Fig. 3(a),
$D_0=1$. However $D_q$ for $q\ne 0$ clearly shows that the spectrum
has multi-fractal properties. For comparison, Fig. 3(e) also shows
the $D_q$ result for a standard kicked-top model with the same
values of $\eta$ and $\alpha$ (i.e., considering an operator
comprising only the first two factors of $F$). The $D_q$ behavior in
the kicked-top case is as trivial as that of a random sequence: it
remains close to unity and slightly decreases with increasing $q$
due to finite-size effects. Based on these results, we conjecture
and invite a formal mathematical proof that the butterfly patterns
found here contain true fractals in the limit of $J\rightarrow
+\infty$.

%**********************************************************************************fig3
\begin{figure}
\vspace{-0.1cm}\epsfig{file=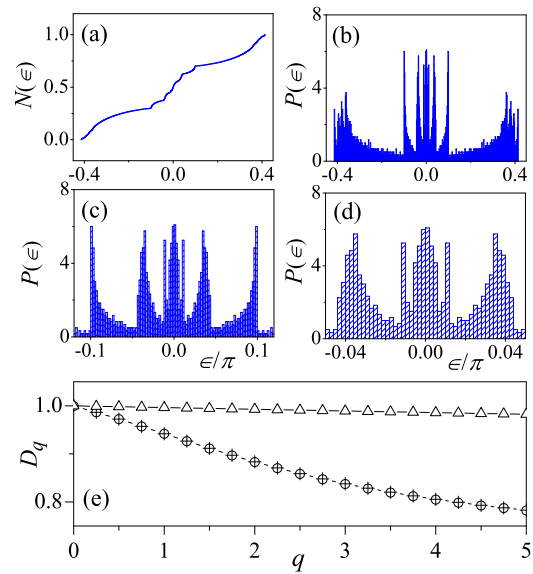,width=7.6cm}
\vspace{-3.1cm}\caption{The cumulative Floquet state density (a) and
the Floquet state density distribution (b)-(d) at different scales,
for $\hbar_{\eta}=(\sqrt{5}-1)\pi/2$, $\alpha/\hbar_{\text{eff}}=1$,
and $J=2999$. Panel (e) shows the generalized fractal dimension
$D_q$. Crosses and circles are for odd-parity and even-parity
states. Triangles represent the result for a standard kicked-top
model.}\label{fig3}
\end{figure}

We next study the level crossings in the butterfly patterns.
Interestingly, the minimal distance in $\hbar_{\eta}$ between two
level crossings is found to decrease sharply  with $J$. So even for
a rather small $J\sim 10$ it is already computationally demanding to
identify all the level crossings. As an example Fig. 4(a) presents
the typical level crossing behavior in the vicinity of a null
eigen-phase for $J=10$.  The Floquet states are seen to cross each
other frequently, between different-parity states and between
same-parity states. Both types of level-crossings are of enormous
interest. For the first type, at a crossing point an arbitrary
superposition of two crossing states of different parities remains
an eigenstate but generally breaks the parity symmetry, thereby
maintaining a nonzero population difference between the two modes
forever \cite{Korsch}. This makes it clear that the first type of
level crossings give rise to the seminal CDT phenomenon
\cite{Hanggi} that has attracted broad experimental interests. Note
that in some regimes of $\hbar_{\eta}$, to the naked eye two curves
of opposite parities in Fig. 4(a) are almost on top of each other,
and as a result many CDT points are found in these regimes. Note
also that CDT-induced population trapping is fundamentally different
from the well-known self-trapping effect on the mean-field level.
Indeed, the CDT effect here depends on $\eta$ and $J$, whereas
mean-field self-trapping is transient and independent of $J$. Now
turning to the second type of level crossings, they come as a
surprise because avoided crossings between same-parity states are
generally anticipated for classically non-integrable systems (see
Fig. 2). The second type of crossings therefore suggest the
uniqueness (e.g., some effective local ``symmetry") of $F$ whose
matrix elements in the $|m\rangle$ representation are
quasi-periodic. Recalling the above-mentioned extreme example where
all levels cross at $\hbar_\eta=2\pi$ for integer $J$, we expect
that special arithmetic properties of $\hbar_{\eta}$ play a key role
in both types of level crossings.

By obtaining all the level crossings with high accuracy for $J\leq
12$, we obtain in Fig. 4(b) that the number of CDT points
contained in the butterfly patterns scales as $J^{3.0}$ and the
number of same-parity crossings scales as $J^{2.7}$.
%In either
%case, the number of crossings divided by the total number of
%levels ($\sim J$) or divided by the total number of level pairs
%($\sim J^2$) diverges as $J\rightarrow +\infty$.
In particular, we
conclude that as $N$ goes to infinity, on average each pair of
Floquet states in a butterfly pattern see infinite CDT points.

%**********************************************************************************fig4
\begin{figure}
\vspace{-0.2cm}\epsfig{file=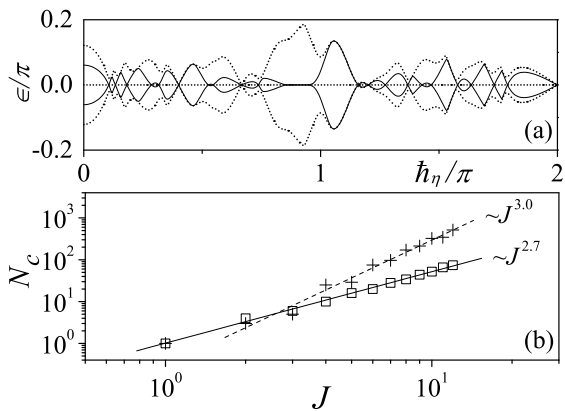,width=8.4cm} \caption{ (a)
Level crossings between three even-parity states (dashed lines) and
two odd-parity states (solid lines), for $J=10$ and
$\alpha/\hbar_{\text{eff}}=1.0$. (b) Number of level crossings
versus $J$, for $\hbar_\eta\in [0,4\pi)$ and
$\alpha/\hbar_{\text{eff}}=1.0$.  The cross (square) symbols are for
crossings between different-parity (same-parity) states and the
fitting suggests a power law scaling $J^{3.0}$ ($J^{2.7}$).}
\label{fig4}
\end{figure}

In the kicked Harper model, the quantization rule varies with the
boundary condition adopted \cite{guarneri} and a compact toroidal
phase space arises only if the Planck constant assumes special
values \cite{Leboeuf}.  A general treatment of the kicked Harper
model leads to a band structure that often complicates the issue. By
contrast, the phase space here is necessarily on a sphere
\cite{Haakebook}, with no arbitrariness in quantization and no band
structure in the spectrum. For these reasons the new butterfly
Floquet spectrum discovered in this work can stimulate more studies
of fractal Floquet spectrum in driven systems. Results here also
suggest that our strategy in generating a butterfly quasi-energy
spectrum, namely, the use of partial cancelation of quasi-random
phases (first advocated in a kicked-rotor system \cite{jiao08}), is
widely applicable. Furthermore,  it is now clear that three quantum
chaos paradigms, i.e., the kicked-top, kicked-Harper, and
kicked-rotor models, are all linked together for the first time,
because all of them can display fractal statistics.

%The
%many level-crossings embedded in the butterfly patterns might also
%motivate formal questions regarding to quantum phase transitions in
%periodically driven systems.

Finally we mention two specific applications.  First, because the
found butterfly spectrum collapses at $\hbar_{\eta}=2\pi$ (or
$g_0=\pi/4$) for integer $J$, one may experimentally determine if
$N$ is even or odd by scanning the dynamics in the neighborhood of
$g_0=\pi/4$. This possibility does not exist in the mean-field
dynamics of a BEC. Similarly, one may study the CDT points to reveal
non-mean-field effects. Second, it is now of great interest, both
experimentally and computationally, to revisit early results of how
a multi-fractal spectrum can be manifested in time-dependent
properties \cite{dynamics}.

Detailed results of this work will be published elsewhere
\cite{jiaogong}. We thank Prof. C.-H. Lai for his encouragement.
J.W. acknowledges support from DSTA of Singapore under agreement of
POD0613356. J.G. is supported by WBS grant Nos.
R-144-050-193-101/133 and R-144-000-195-123.

%Possible application:

%Slow decay of temporal correlations in quantum systems with Cantor
%spectra, (Ketzmerick),  PRL69, 695.  (1992).  What if we study the
%double kicked top, can we check these predictions? These will become
%possible nowdays! $D_0$, and $D_2$ dimensions will be very very
%important!  Phase imaging, or the population imbalance, phase
%relationship etc...  What about the fidelity? Will the fidelity
%decay exponent connected with the fractal dimension? This will also
%be very important.

%Entanglement generation, detection in flufucations in external
%parameters because of sensitity of the spectrum.  Probably a great
%system to detect entanglement generation etc..

\end{document}